\documentclass[12pt]{iopart}

\usepackage{young}
\usepackage{color}
\usepackage{pstricks}
\usepackage{graphicx}
\begin{document}

\title{Chiral Symmetry Breaking in Graphene}

\author{Gordon W. Semenoff}

\address{Department of Physics and Astronomy, University of British Columbia, 6224 Agricultural Road, Vancouver, British Columbia, Canada V6T 1Z1}
\begin{abstract}
The question of whether the Coulomb interaction is strong enough
to break the sublattice symmetry of un-doped graphene is discussed.
We formulate a strong coupling expansion where the ground state of the
Coulomb Hamiltonian is found exactly and the kinetic hopping Hamiltonian
is treated as a perturbation.   We argue that many of the properties of the
resulting system would be shared by graphene with a Hubbard model interaction.
In particular, the best candidate sublattice symmetry breaking ground state is
an antiferromagnetic Mott insulator.  We discuss the results of some
numerical simulations
which indicate that the Coulomb interaction is indeed subcritical.
We also point out the curious fact that, if the electron did not have spin degeneracy,
the tendency to break chiral symmetry would be much greater and even relatively weak Coulomb
interactions would likely gap the spectrum.
\end{abstract}

\maketitle

\section{Introduction}

Some of the original work in the prehistory of graphene  \cite{Semenoff:1984dq} was motivated by an attempt to find
an analog in condensed matter physics of some very nice structures which, in the early 1980's,
had emerged in the study of relativistic quantum field theories in 3 space-time dimensions.  These
included the appearance of Chern-Simons terms and topological mass for gauge fields \cite{Deser:1981wh}
 \cite{Deser:1982vy}, the parity
anomaly \cite{Niemi} \cite{Redlich:1983dv} \cite{Redlich:1983kn}
and the use of the index theorem
\cite{Niemi} \cite{Jackiw:1984ji}-\cite{Niemi:1984vz} to learn about features of the spectrum of the Dirac Hamiltonian.
When graphene was discovered and made readily available in the laboratory \cite{novoselov:2004}, as luck would have it,  many
of its features turn out to be well described by free 2+1-dimensional relativistic
fermions with an emergent U(4) symmetry.  The index theorem plays a role.  It determines the degeneracy
of the states at zero energy, when the electrons are exposed to an external magnetic field.
These   states  are responsible for the anomalous integer quantum Hall effect
\cite{graphenehalleffect}-\cite{antonio}, which had an important historical role in that it was
the experimental result which drew the attention of the larger physics community as
many considered it the smoking gun of relativistic Dirac electrons
in graphene.  The other effects, the parity anomaly and induced Chern-Simons terms are less
directly visible in graphene and have had to await the advent of topological insulators \cite{kane}-\cite{kane3} to come
into their own.
 This is by now well-worn ground, as evidenced by the excellent presentations at this symposium.

 What
I want to talk about today is another important subject, one of dynamics, the effects of the strong Coulomb interaction
in graphene.   This dynamics also has an analog in relativistic quantum field theory, and in elementary particle physics, in
the study of chiral symmetry breaking.
The phenomenon of spontaneous chiral symmetry breaking is one of the cornerstones
of our current understanding of the strong nuclear interactions.  The
approximate chiral symmetry of almost massless quarks is spontaneously
broken.  Quarks gain mass and the Goldstone bosons are pions which, because the symmetry
is not exact, are light but not massless.  It is a general view that this
spontaneous symmetry breaking is driven by strong gauge field mediated
interactions, in the case of quantum chromodynamics, the exchange of gluons.  It proceeds through the formation of
a mass operator
condensate, $\left<\bar\psi\psi\right>$.   This condensate breaks chiral symmetry and fermion masses are the result.
This phenomenon is important for the strong interactions, but it could also be more far-reaching, as it is part of the
circle of ideas behind some extensions of the standard model, technicolor models being an example.

Being a strong coupling phenomenon, chiral symmetry breaking is notoriously
difficult to understand in a quantitative way.  The standard quantum field theory
tool of perturbation theory is not
reliable in the strong coupling regime.
One interesting way to gain intuition about chiral symmetry breaking
has been to study the analogous phenomenon is simpler models, like
2+1-dimensional electrodynamics.

This toy model of chiral symmetry breaking: \cite{Jackiw:1980kv}-\cite{Appelquist:2004ib}
is a  2+1-dimensional quantum field theory with U(N) symmetry:
\begin{equation}
{\cal L}(x) = \sum_{a=1}^N i\bar\psi_a(x) \gamma^\mu D_\mu\psi_a(x)-
\frac{N}{4g^2}F_{\mu\nu}F^{\mu\nu}
\label{chiralsymmetrybreakingmodelar}
\end{equation}
The minimal representation of the Dirac matrix algebra in three spacetime dimensions is two --
the fermions are a doublet spinor representation of the $SO(2,1)$ Lorentz symmetry.
Chirality symmetry in three spacetime dimensions is intimately tied to symmetry under parity and time-reversal.
Generally, in any dimension, massless fermions
have more symmetries than massive fermions.  It is not possible to find a mass term $\sim\bar\psi\psi$
which has all of the symmetries of the kinetic term in the fermion Lagrangian,
$i\bar\psi\gamma^\mu\partial_\mu\psi$.
The kinetic term in the three dimensional fermion action are unaffected 
if we implement a parity transformation by making the replacement
\begin{equation}\label{parity}
\psi(x,y,t)\to \gamma^1\psi(-x,y,t)
\end{equation}
and change the integration variables accordingly.    The kinetic term in the action,
\begin{equation}\int dtdxdy~\bar\psi i\gamma^\mu\partial_\mu\psi\end{equation}
is invariant.  However, a mass term, which is of the form
\begin{equation}
\int dtdxdy~m\psi^\dagger\gamma^0\psi
\end{equation}
changes sign under parity.  One can formulate a parity invariant mass term if there is more than one species of fermion.
For example, if
there were two species and they had opposite signs of mass, so their mass terms were of the form
\begin{equation}
m\left[\bar\psi_1\psi_1 - \bar\psi_2\psi_2\right]
\label{parityinvariantmassterm}
\end{equation}
we could define a parity transformation by augmenting the transformation in
(\ref{parity}) by an exchange of the two species,
$$
\psi_1(x,y,t)\to \gamma^1\psi_2(-x,y,t)~~,~~
\psi_2(x,y,t)\to \gamma^1\psi_1(-x,y,t)
$$
The mass term (\ref{parityinvariantmassterm}) is invariant under this transformation,
so parity is restored.  On the other hand, the kinetic
Lagrangian
$$
\bar\psi_1i\gamma^\mu\partial_\mu\psi_1 + \bar\psi_2i\gamma^\mu\partial_\mu\psi_2
$$
has a  symmetry of exchanging $\psi_1\leftrightarrow\psi_2$ which the mass term (\ref{parityinvariantmassterm}) doesn't have.
This symmetry would actually be a larger continuous internal symmetry $U(2)$ or $O(2)$,
depending on whether we are discussing complex Dirac fermions
or charge-self-conjugate Majorana fermions.

If, on the other hand, we introduced the $\psi_1\leftrightarrow\psi_2$  symmetric mass term,
$$
M\left[\bar\psi_1\psi_1+\bar\psi_2\psi_2\right]
$$
it would break parity.  There is no way to introduce a fermion mass without breaking either parity or
some of the internal symmetry.

The above argument applies equally well if $\psi$ is a complex Dirac fermion or a charge-self-conjugate Majorana
fermion.
A complex fermion can be written as two Majorana fermions.  Let us choose a specific representation
for the Dirac matrices,
$$
\gamma^0=\sigma^2~~,~~
\gamma^1=i\sigma^3~~,~~
\gamma^2=i\sigma^1
$$
where $\sigma^a$ are Pauli matrices. With this convention, the Dirac equation
$$
\left(i\gamma^\mu\partial_\mu + m\right)\psi=0
$$
is real, so it is satisfied by both $\psi$ and $\psi^*$, the complex conjugate.
Then the complex fermion can be written as
$$
\psi=\frac{1}{\sqrt{2}}\left( \psi_R+i\psi_I\right)
$$ where $\psi_R=\psi_R^*$ and $\psi_I=\psi_I^*$
and we can write the action for a complex fermion as  the sum of two actions, one for
the real and one for the imaginary part,
$$
\frac{1}{2}\bar\psi_R i\gamma^\mu\partial_\mu\psi_R
+\frac{1}{2}\bar\psi_I i\gamma^\mu\partial_\mu\psi_I
$$
We could introduce a parity invariant mass term for this system,
$$
\frac{m}{2}\left[\bar\psi_R\psi_R-\bar\psi_I\psi_I\right]
$$
Note that this term can be non-zero, consistent  with Fermi statistics.
If $\psi_R=\left[\begin{matrix}{u\cr v\cr}\end{matrix}\right]$ then
$\frac{m}{2}\bar\psi_R\psi_R~=~ m~uv$ if $u$ and $v$ anticommute with each other.
In terms of the original complex fermion, this mass term is
$$
\frac{m}{2}\left[ \psi^t\gamma^0\psi + \psi^\dagger\gamma^0\psi^*\right]
$$
which breaks the phase symmetry of the complex fermion.
Such a mass term could appear in a superconducting
state, for example, if superconductivity were induced in graphene using the proximity
effect, it would be possible to have a relativistic mass term without breaking parity.
In the following, we are going to always assume that the phase symmetry is not broken.
In that case, we can limit the discussion to complex fermions.  Then, if this phase
symmetry is not broken, the possible mass terms for complex fermions must either
break parity or what is generally an internal unitary symmetry.  For example, graphene
has emergent $U(4)$ symmetry.  The phase symmetry corresponding to charge conservation, and which
is gauge to get electromagnetic interactions is the $U(1)$ subgroup of $U(4)$.  We are assuming
that this subgroup survives in any chiral symmetry breaking scheme.  The sort of symmetry breaking
that we will consider will generally be $U(4)\to U(2)\times U(2)$ or $U(4)\to U(1)^4$.

The quantum field theory (\ref{chiralsymmetrybreakingmodelar}) has a dimensional coupling constant, $g^2$.
It is super-renormalizable in that all but a finite number of Feynman diagrams
are ultraviolet divergent.  However, perturbation theory in   $g^2$
 leads to severe infrared divergences \cite{Jackiw:1980kv}.  Moreover, the low energy and momentum limit of
 the theory is strongly coupled.  An alternative, renormalizable expansion can
 be carried out using the dimensionless parameter $\frac{1}{N}$ \cite{Pisarski}.   The theory is solvable in the large $N$ limit,
 that is, when the number of fermion
species $N$ is taken to infinity, and there is a systematic expansion in $\frac{1}{N}$.
However, in the large $N$ limit, the  interaction is of order $\frac{1}{N}$ and it is
too weak  to break chiral symmetry.
It is thought that, as $N$ is lowered to smaller values, and the interaction gets
stronger, it eventually reaches a critical value after which chiral symmetry
is broken.  Estimates of critical $N$ vary from 1.5-4.5, the spread being a symptom of
the lack of accuracy of the large $N$ expansion when $N$ is small. To implement the large $N$ expansion, the
fermions are integrated out to produce an effective action
\begin{eqnarray}\label{chiralsymmetrybreakingmodel}
S_{\rm eff}&&= -\frac{N}{4g^2}\int d^3x~F_{\mu\nu}F^{\mu\nu}-iN{\rm Tr}\ln\left[\gamma\cdot(i\partial+A)\right] \nonumber \\
&&\approx -\frac{N}{4}\int d^3x F_{\mu\nu}\left( \frac{1}{g^2}+ \frac{1}{4\sqrt{-\partial^2}}\right)F^{\mu\nu}+~{\rm interactions}
\end{eqnarray}
In the remaining perturbation theory, $g^2$ acts as a cutoff in that is softens the ultraviolet behavior of the
effective photon propagator $\frac{4}{p}\to \frac{ g^2 }{g^2 +4p }\frac{4}{p}$.  To study the behavior of the theory in the infrared, that is
where characteristic momenta $p^2<<g^4$, we could put $g^2$ to infinity and replace it with another cutoff. The resulting
effective theory has a certain resemblance to graphene with a coulomb interaction, where $N=4$, as we shall outline below.

In graphene, which is a 2-dimensional hexagonal array of carbon atoms,
nature has given us an emergent example of a system of interacting
 relativistic fermions with   $U(4)$
symmetry.  Its low energy ($<1ev$) dynamics are described by the continuum field theory with
action
\begin{eqnarray}
S=\int d^3x~\sum_{k=1}^4 \bar\psi_k\left[\gamma^t(i\partial_t-A_t)+v_F\vec\gamma\cdot(i\vec\nabla-\vec A)\right]\psi_k
\nonumber \\~~~~~~~~~~~~~~~~~~~~~~~~~~~~~~~~
  -\frac{\epsilon}{4e^2 }\int d^3x~F_{ab}\frac{1}{2\sqrt{-\partial^2}}F^{ab}
\label{graphene}
\end{eqnarray}
 The non-local nature of the gauge field action is due to the fact that the photon
 which mediates the interaction between electrons
 propagates out of the graphene plane, in 3+1-dimensional space-time.   This  results in
 the non-local last term in the 2+1-dimensional action (\ref{graphene}). The dimensionless constant $e$ is
 the charge of the electron and $\epsilon$ is the dielectric constant which could differ from that of the vacuum if
 graphene is immersed in another material, or attached to a substrate.   In the following, we shall consider
 this theory at zero density, or the so-called charge neutral point only.

 The parameter $v_F$ is the velocity of the massless
 electron in graphene.   It is different from the vacuum speed of light at which the photon propagates (and which we have set
 to one ($\hbar=1=c$)), in fact $\frac{v_F}{c}\sim \frac{1}{300}$.  This makes (\ref{graphene}) a non-relativistic field theory, unlike the model (\ref{chiralsymmetrybreakingmodel}) which we discussed above. The Coulomb interaction is to a first
 approximation instantaneous and the magnetic interactions are suppressed by factors of $\frac{v_F}{c}$. Another essential difference from (\ref{chiralsymmetrybreakingmodel}) is the coefficient of the nonlocal term in the action.  In (\ref{chiralsymmetrybreakingmodel}) it was fixed by the number of fermions that were
  integrated out, whereas in (\ref{graphene}) it is determined by the dielectric constant of the medium
  in which graphene is immersed. 

Although the graphene model (\ref{graphene}) is nonrelativistic, our discussion of chiral symmetry breaking applies to it
as well as the relativistic theory.  A mass term for the fermion would break either parity symmetry or some of the emergent $U(4)$ symmetry.  In fact, at the lattice level, the relevant parts of the U(4) symmetry are replaced by sublattice symmetry -
the symmetry of the graphene Hamiltonian under interchanging the two triangular lattices that make up the honeycomb lattice.
Breaking this symmetry is intimately related with forming a gap in the fermion spectrum \cite{Semenoff:1984dq}.

\begin{figure}
 ~~~~~~~~~~~~~~~~~~~~~~~~~~~~~~~~~~~~~~~~~~~~~~~~~\includegraphics[scale=.8]{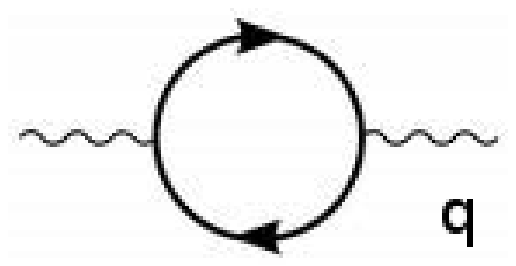}\\
\begin{caption} {A Feynman diagram which contributes to the polarization tensor.  The
graphene fine structure constant quoted in eq.~(\ref{fsc}) emerges as a coefficient.  \label{vacuumpolarization}
}\end{caption}
\end{figure}

The nonrelativistic field theory with action (\ref{graphene}) is a renormalizable quantum field theory, in the sense that, in perturbation theory in
 its dimensionless coupling constant $e^2$, no new counterterms have to be introduced to cancel ultraviolet divergences \cite{voz}-\cite{voz8}.  This
 has been confirmed explicitly up to order two loops.  Renormalizing perturbation theory requires a logarithmically divergent counterterm for $v_F$, which then becomes a scale-dependent parameter, running to larger values in the infrared limit.
The perturbative beta function vanishes at the
 Lorentz invariant limit, $v_F=c$.  It is conjectured that the coupling constant $e^2$ remains a tuneable parameter
 so that the theory has a fixed line where it would be a conformal field theory with a tuneable constant $e^2$.
 However, if one uses the one-loop beta function, and the known value of $v_F$ at the
 lattice scale as a boundary condition, $v_F$ runs so slowly that to even get
 to the order of $c/10$ requires wavelengths of the magnitude of meters,
 larger than  the length scales achievable in experiments. Moreover, the graphene fine structure constant
which controls loop corrections in perturbation theory, is large.\begin{equation}\label{fsc}\alpha_g\equiv \frac{ e^2}{4\pi\hbar
v_F}= \frac{ e^2}{4\pi\epsilon\hbar c}\frac{c}{
v_F}\approx\frac{300}{137}\end{equation}
This is the constant which occurs in loop integrals for quantities such as the polarization tensor given by the Feynman
diagram in figure \ref{vacuumpolarization}.
If the effective coupling constant is really this large, the accuracy of perturbative computations is questionable, at best.

The central question which we shall address in the following is, given that the Coulomb interaction is strongly coupled,
whether the interaction is strong enough to break chiral symmetry.   Part of the question is to ask whether, if the coupling
were variable, there is a critical value of the coupling where  a quantum phase transition occurs, particularly to a phase
 where chiral symmetry is broken.  The second question is as to whether the physical parameters of graphene put it
 in or close to this range.  The current experimental status of graphene, where there is no evidence of spontaneous gap
 formation, at least in the absence of strong magnetic fields \cite{cat1}-\cite{cat6},
 suggests that the answer to the second question is no.
 If there would be a quantum phase transition, the coupling
 in real graphene is apparently not strong enough.  The fact that it could be close to being
 strong enough is also tantalizing, as some
  mechanical or other physical deformation of graphene could then make it closer, and drive it to the phase transition.
  We will spend the next Section on an attempt to understand why the strong coupling in graphene might be strong enough,
  and why it might be close, but we do not as yet have a quantitative estimate of how close.

 We should mention that this idea has been pursued in some lattice simulations \cite{Armour:2009vj}-\cite{drut2010B}.
 It has also been addressed using a lattice strong coupling expansion \cite{araki}.  These studies do not use the graphene
 lattice, but a square lattice with magnetic flux to produce relativistic fermions.  They seem to answer the question as to
 whether chiral symmetry breaking would exist in the affirmative.  One uncertainty that remains is whether the same results
  would hold for a honeycomb lattice.   One might expect universal features such as critical exponents to be predictable, however the nonuniversal details like the value of critical couplings could differ.

  Finally, there is a large $N$ expansion
  which indicates that sufficiently strong contact interactions could drive chiral symmetry breaking.  This fact has been
  known for a long time in the particle physics literature and it is exploited for the physics of graphene in references
\cite{herbut} \cite{herbut2} \cite{Son:2007ja}-\cite{voz8}. In references \cite{herbut} \cite{herbut2}, they  find that the local on-site interactions are crucial and that the Coulomb interaction
makes only a weak modification of the critical behavior which is essentially driven by the point interactions do.  This is not a lot different from what we shall find.

 Before we proceed, we should mention another idea which has been pursued recently, the use of AdS/CFT holography to study
 the strong coupling limit of field theories which could be regarded as deformations of graphene. Interestingly, using
 holography, it is possible to construct two different scenarios\cite{Kutasov:2011fr} \cite{Davis:2011gi}.  In one \cite{Kutasov:2011fr}, there is a strong coupling fixed point for the analog of $e^2$ where chiral symmetry is broken if $e^2$ is sufficiently large, and the mass gap is small only if $e^2$ is tuned
 to be sufficiently close to this fixed point.  In the other construction \cite{Davis:2011gi}, $e^2$ is tuneable all the way to infinity and there is no phase transition.  Instead the theory is a nontrivial conformal field theory and chiral symmetry is
 only broken after turning on
a dangerous relevant operator, related to the mass operator. Interestingly, the first scenario, where the field theory inhabits a planar defect in a single 3+1-dimensional gauge theory, and where there is a chiral phase
transition, bears a certain resemblance to suspended graphene, whereas the second scenario, where the field theory inhabits a planar boundary between two different gauge theories,  has a similarity to graphene on a substrate.

\section{Strong coupling expansion}

If the graphene fine structure constant is really as large as the
estimate in (\ref{fsc}) perturbation theory is of no use in analyzing
the effects of the Coulomb interaction. The alternative of a strong coupling expansion would
be the more reasonable approach.   Strong coupling expansions of relativistic gauge field
theories, once they are regulated by putting them on a lattice, give a nice
qualitative picture of confinement.  However, they are far from the continuum limit
and are notoriously non-universal, particularly for questions involving chiral symmetry breaking,
where different definitions of the lattice theory which have the same naive continuum limit, can have
wildly different results.  However, graphene comes from a lattice to begin with. One could begin
with a reasonable estimate  of the lattice Hamiltonian of graphene as the starting point
for a strong coupling expansion.

We will assume that graphene is modeled by the following Hamiltonian,
\begin{equation}\label{fullhamiltonian}  H =H_t+H_e
\end{equation}
where the hopping term  is
\begin{equation}\label{hopping}
H_t=
t\sum_{A,i,\sigma}\left(\psi^\dagger_{\sigma,A+{\bf s}_i}\psi_{\sigma, A} +\psi^\dagger_{\sigma, A}\psi_{\sigma,A+  {\bf s}_i}\right)
\end{equation}
and the Coulomb interaction  is
\begin{equation}\label{coulomb}
H_e=\frac{e^2}{8\pi\epsilon a}~\sum_n   u_0~\rho_n^2
 + \frac{e^2}{8\pi\epsilon a}~\sum_{n\neq n'}
\rho_n
\frac{1}{  |n-n'|}
\rho_{n'}
\end{equation}
The electron creation and annihilation operators are denoted by $ \psi_{\sigma,n}^\dagger$ and $\psi_{\sigma,n}$.
They
 have spin label $\sigma$ with two spin states $\sigma=\uparrow$ and
 $\sigma=\downarrow$ and site label $n$ which can be on either the $A$ or $B$ sublattice.
 The non-vanishing
 anti-commutators  are \begin{equation}\{\psi_{\sigma,n},\psi_{\sigma',n'}^\dagger\}=\delta_{nn'}\delta_{\sigma\sigma'}
 \end{equation}
 The space of quantum states is a Hilbert space constructed by cyclic action of creation operators on the empty ``vacuum''
 state,   $ \left|0\right>$, which obeys $\psi_{\sigma,n}\left|0\right>=0$ for all values of the
 labels $\sigma$ and  $n$.\footnote{We are ignoring subtleties of exactly how this Hilbert space would be defined
   in a system with infinite volume.}  We will consider the case where the graphene sample is neutral, that is, half
 of the available electronic states are occupied.  The electron density operator in (\ref{coulomb}) is given by
 \begin{equation}
 \rho_n = \psi^\dagger_{\uparrow,n}\psi_{\uparrow,n}+\psi^\dagger_{\downarrow,n}\psi_{\downarrow,n}-1
 \label{density}\end{equation}
The ``-1'' in the density represents the positive ion which occupies each site of the lattice.
  The spectrum of the electron charge operator $\psi^\dagger_n\psi_n$
is $0,1,1,2$ and the spectrum of $\rho_n$ is $(-1,0,0,1 $),
according to whether the site is unoccupied, singly occupied or doubly occupied, respectively.
Double occupation is allowed by Fermi statistics because there are two spin states.

\begin{figure}
 ~~~~~~~~~~~~~~~~~~~~\includegraphics[scale=.4]{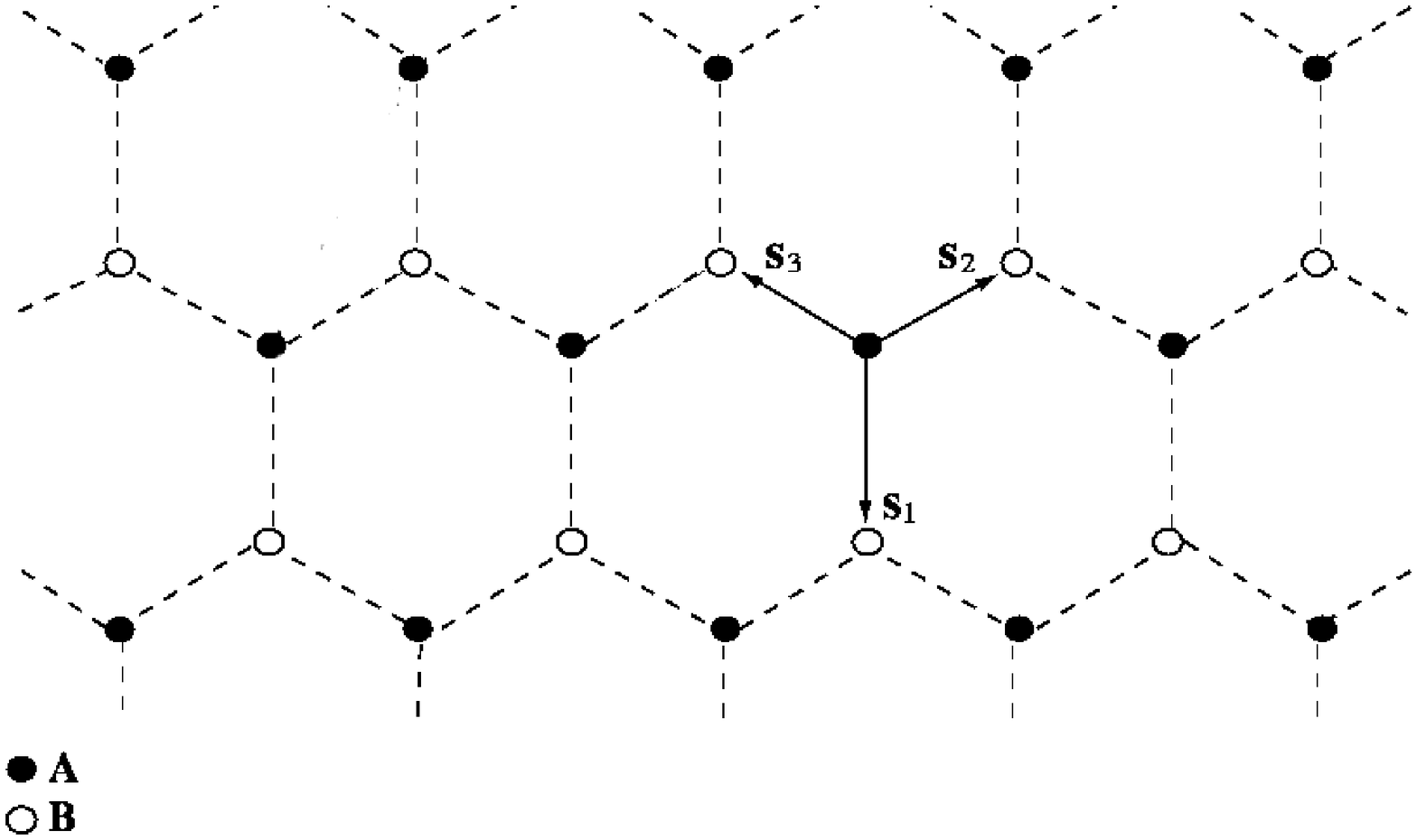}\\
\begin{caption} {\label{sublattices}
The honeycomb graphene lattice with $A$
and $B$ sublattices depicted by black and white dots, respectively.
The sublattices are connected by ${\bf s}_i$.}\end{caption}
\end{figure}

 $H_t$ describes the energy due to tunneling between tight binding states on adjacent lattice sites. $A$ labels a point
 on sublattice $A$ and the points $A+{\bf s}_i$ are on sublattice $B$. The lattice is depicted in figure \ref{sublattices}.
  It is the continuum limit of $H_t$ which
 produces the Dirac Hamiltonian which is used for the description of electron
 dynamics close to the fermi level in  graphene.
 The hopping
 amplitude in $H_t$ has the approximate magnitude $t\sim 2.7ev$.

The Coulomb interaction is described by $H_e$. In $H_e$, the site labels
 $n$ and $n'$ are summed over both the A and B sublattices and we are using units $\hbar=1=c$.
 The constant $a$ in the last term is the lattice constant of graphene,
$a\sim 2.461${\AA}. We are also
using a dielectric constant $\epsilon$.  With the dielectric constant of the vacuum,
$e^2/4\pi\epsilon_0 a \approx 5.85ev$.  If one used the dielectric constant of silicon dioxide,
$\epsilon\approx 3.9\epsilon_0$, this characteristic energy would be lower.

We have also modeled the on-site Coulomb interaction
with a term which penalizes the state at a given site for having charge, with the
same penalty for either positive or negative charge.  It could easily be modified
to make these different.  $u_0$ parameterizes
this on-site Coulomb interaction.  The magnitude of $\frac{e^2}{4\pi\epsilon a}u_0$ is in the range $\sim 5-10ev$.
\footnote{In a recent paper (\cite{katsetal}) $\frac{e^2}{4\pi\epsilon a}u_0$ is estimated to be $9.3ev$.}
This should be compared with the kinetic, hopping energy is $\sim 2.7ev$.

The presence of $u_0$ is essential for us.  It is important to make
the Coulomb interaction kernel positive.  To see this, consider the Coulomb energy of an electron and a hole
separated by one lattice spacing,
\begin{equation}
E_{\rm eh}= \frac{e^2}{4\pi\epsilon a}(-1+u_0)
\end{equation}
The first term with ``-1'' is the interaction between the electron and hole.  At any finite separation, this interaction energy
is negative and it goes to zero if they are infinitely separated.  The other term, with $u_0$, is the on-site (self-)energy of
the electron and the hole. Physically, it is quite plausible that it takes a finite positive amount of energy to
separate an electron and a hole by one lattice spacing.  This will only be the case in our model if $u_0>1$. We
will assume that this is the case.

The sort of gap generation that we will be looking for is one which breaks the lattice translation symmetry spontaneously.
An order parameter for this symmetry breaking is the expectation value of the following operator
\begin{equation}\label{massterm}
H_m = \left(\sum_{n\in A}-\sum_{n\in B}\right)\left[ \mu_0\psi^\dagger_{\sigma,n}\psi_{\sigma,n}+\vec\mu\cdot\psi^\dagger_{\sigma,n}\vec\sigma_{\sigma\sigma'}\psi_{\sigma',n}
\right]
\end{equation}
It was shown in reference \cite{Semenoff:1984dq} that adding $H_m$ to $H_t$ in (\ref{hopping})  would lead a mass term for Dirac Hamiltonian (\ref{graphene}) which is obtained in
the continuum limit of $H_t$ when the system is charge neutral.
This sort of mass term is invariant under time reversal and parity, but it breaks some of the emergent U(4) symmetry of the continuum limit.   If both parameters $\mu_0$ and $|\vec\mu|$ are nonzero, the symmetry breaking
pattern is $U(4)\to U(1)^4$.  If either $\mu_0$ or $|\vec\mu|$ is zero, but the other nonzero, the pattern is
$U(4)\to U(2)\times U(2)$.  For the most part, it is the latter symmetry breaking pattern that we shall
discuss in the following.  At this point, we should note that there are other kinds of symmetry breaking which are possible.
These have to do with distortions of the graphene lattice which lead to bond order, like the Kekule distortion.  That
sort of symmetry breaking is analogous to the Pierls instability of one-dimensional tight binding models and it an
important possibility for graphene and some of the interesting consequences have been pursued in a number of works \cite{JackiwPi}-\cite{Chamon:2007hx}.

 Our goal is to find the ground state of the full Hamiltonian in (\ref{fullhamiltonian}).  We are
 not able to find an exact solution.  We therefore have to resort to an approximation. Since, as we
 have argued above, the Coulomb energy is typically larger than the kinetic energy of electrons, we
 will begin by seeking a ground state of the Coulomb Hamiltonian, $H_e$.

\begin{figure}
~~~~~~~~\includegraphics[scale=1]{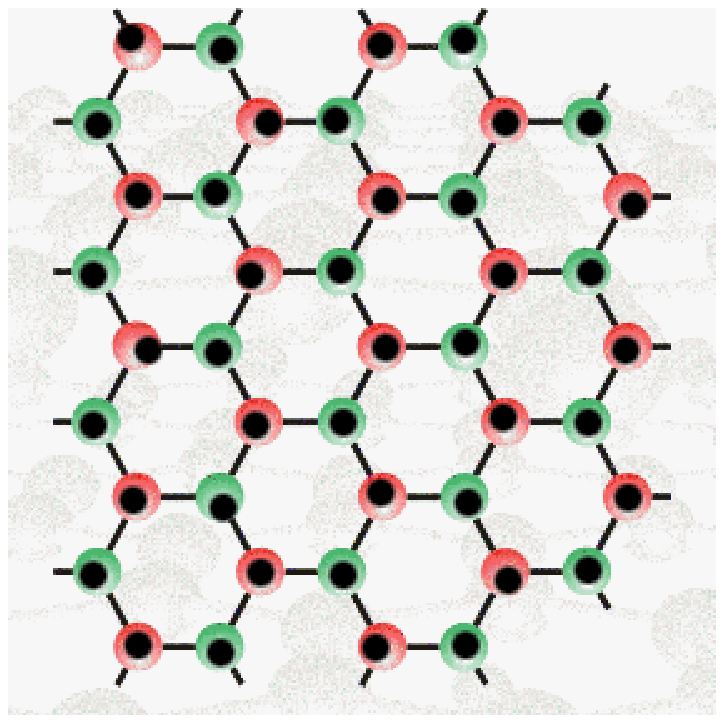} \\
\begin{caption}{\label{graphenemottinslator}The lowest energy state of the Coulomb interaction has
one electron localized at each site.  This is a highly degenerate state as the
spins of the electrons are arbitrary.}\end{caption}
\end{figure}

 To do this, we first observe that the charge density operators are Hermitian and, at different sites,
 they commute with each other,
  \begin{equation}
  \left[\rho_n,\rho_{n'}\right]=0
  \end{equation}
  Thus, the charge densities, and therefore also the Coulomb Hamiltonian $H_e$
 can be simultaneously diagonal.
 This fact effectively makes finding the
 lowest energy state of $H_e$ a classical problem - we need only find that classical   distribution
 of unit point charges  which minimizes the Coulomb energy.
 If we consider the fourier transform of the charge density
 \begin{equation}
 \rho_n = \int_{\Omega_B} \frac{d^2k}{\sqrt{\Omega_B}}~e^{ik\cdot n}~\hat\rho(k)
 \end{equation}
 where $\Omega_B$ is the Brillouin zone of the triangular $A$ sublattice.
 The Coulomb energy is diagonal in charge densities,
 \begin{equation}
 H_e= \frac{e^2}{4\pi\epsilon a}\int d^2k \left|\hat\rho(k)\right|^2\left( \frac{u_0}{2}+ \sum_{n\neq 0,n\in A}e^{ik\cdot n}
  \left[\frac{1}{ |n |}+\frac{\cos{k\cdot{\bf s}_1}}{|n+{\bf s}_1|}\right]\right)
 \end{equation}
 The absolute minimum is where $\hat \rho(k)=0$.   This is compatible with the precisely half-filled lattice
 that we are considering since the constraint of half-filling is $\hat \rho(0)=0$. This is the completely neutral state. This neutral state
 has one electron occupying each lattice site, as in figure \ref{graphenemottinslator}.

 However, we are not finished yet.
 This is a highly degenerate state, as each electron have either spin polarization, up or down. The state itself is
  \begin{equation}
  \left|\sigma_n\right>~=~\prod_n \psi_{\sigma_n,n}\left|0\right>
  \end{equation}
  where $\sigma_n$ is the spin orientation at site $n$.  The degeneracy of this state is $2^V$ where $V$ is the total
  number of sites.  All of these states are insulators. The linear combination of them which becomes
   is the preferred ground state will also be an insulator.  The spectrum of the electron has a mass gap.

  As usual, this degeneracy must be resolved in perturbation theory by including perturbations from the kinetic Hamiltonian $H_t$.
 The states are mixed at second order in perturbation theory, where $H_t$ can implement transport from a site to a neighbor and then back again.   When the process is allowed, second order perturbation theory always lowers the ground state energy.
 Here, it is allowed when the spin orientation of neighbors is opposite.  The result is that the lower energy state is
 an antiferromagnet.   To see this in technical terms, the effective Hamiltonian for the degenerate ground states is
 \begin{eqnarray}
 H_{\rm eff}=  -H_t\frac{1}{H_e}H_t \nonumber \\
 = \frac{-t^2}{\frac{e^2}{4\pi\epsilon a}(u_0-1)}\sum_{A,{\bf s}_i,\sigma }
 \left(
 \psi^\dagger_{\sigma,A+{\bf s}_i}\psi_{\sigma, A} \psi^\dagger_{\sigma, A}\psi_{\sigma,A+  {\bf s}_i}
  +  \psi^\dagger_{\sigma, A}\psi_{\sigma,A+  {\bf s}_i}\psi^\dagger_{\sigma,A+{\bf s}_i}\psi_{\sigma, A} \right)
 \end{eqnarray}
 Using the sum rule for Pauli matrices
 \begin{equation}
 \vec \sigma_{\sigma\sigma'}\cdot\vec \sigma_{\sigma''\sigma'''}=2 \delta_{\sigma\sigma'''}\delta_{\sigma'\sigma''}
 -\delta_{\sigma\sigma'}\delta_{\sigma''\sigma'''}
 \end{equation}
 we can write the effective Hamiltonian as
 \begin{equation}\label{antiferromagnet}
 H_{\rm eff}= \frac{t^2}{\frac{e^2}{4\pi\epsilon a}(u_0-1)}\sum_{A,{\bf s}_i}
 \psi^\dagger_{ A+{\bf s}_i}\vec\sigma\psi_{ A+  {\bf s}_i}\cdot \psi^\dagger_{  A}\vec\sigma\psi_{  A}
 \end{equation}
   which is the Hamiltonian of the quantum Heisenberg anti-ferromagnet with nearest neighbor coupling. We have
   dropped terms containing the charge density, as all states that this Hamiltonian would act on are eigenstates
   of charge density with zero eigenvalue.  The magnitude of the coefficient in front of the Hamiltonian is $1ev$, which
   sets the scale of both the ground state energy and the energy of excitations of the antiferromagnet.
   The ground state of the Hamiltonian in (\ref{antiferromagnet}) is known to be
   an anti-ferromagnetically ordered state which
   breaks the sublattice symmetry.  The classical
   approximation to the ground state of the antiferromagnet with order parameter oriented along the z-axis is
   \begin{equation}
   \left|{\rm afm}\right>=\prod_A \psi^\dagger_{\uparrow,A}  \prod_B\psi^\dagger_{\downarrow,B} \left| 0\right>
   \end{equation}
In $H_m$ in equation (\ref{massterm}), the second term would have non-zero
   expectation value.  The magnitude of this expectation value is somewhat smaller than
   the coefficient of the effective
   Hamiltonian in (\ref{antiferromagnet}).

\begin{figure}
~~~~~~~~\includegraphics[scale=1]{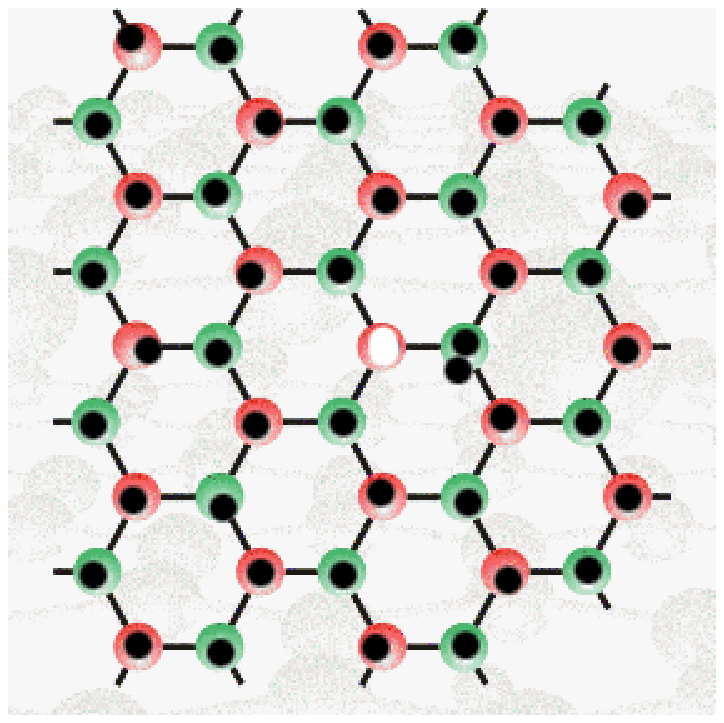} \\
\begin{caption}{\label{elementaryexcitation}
The elementary excitation of the lattice with one electron at each site.
The energy of this state is denoted $U$ and is of the order of $10ev$.}\end{caption}
\end{figure}

The Coulomb ground state that we have come up with is identical to the ground state
of the half-filled Hubbard model.  In addition, the lowest energy charged excitation, depicted
in figure \ref{elementaryexcitation} is also identical to the lowest energy charged
excitation of the half-filled Hubbard model.
To describe it and its very low energy
dynamics, it is tempting to replace the Coulomb Hamiltonian by the Hubbard model whose Hamiltonian is
\begin{equation} H =
t\sum_{A,i}\left(\psi^\dagger_{A+{\bf s}_i}\psi_A+\psi^\dagger_A\psi_{A+{\bf s}_i}\right)
+ \frac{U}{2}\sum_{n\in
A,B}\left(\sum_{\sigma=\uparrow\downarrow}\psi^\dagger_{\sigma n}\psi_{\sigma n}-1\right)^2
\label{hubbard}
\end{equation}
where $U=\frac{e^2}{4\pi\epsilon a}[u_0-1]$ is the energy of the
state in figure \ref{elementaryexcitation}. The Hubbard model is short-ranged.  It simply
penalizes non-single occupation of sites, in the large $U$ limit, projecting onto singly occupied sites.
Note that the Coulomb Hamiltonian is a little stronger than  Hubbard in that, for Hubbard
the energy of the lowest excitation -- the electron-hole pair -- is independent of the
distance between them, whereas for Coulomb, they have opposite charge and would still attract and would
tend to fall  together.  This should make the ground state that we are discussing more stable
for the Coulomb Hamiltonian than for the Hubbard model.  It might therefore have a higher critical coupling.

The Hubbard model (\ref{hubbard}) and the Heisenberg antiferromagnet
on a honeycomb lattice have been studied both by analytic and numerical techniques \cite{sorella}-\cite{wang}.
It is known
that in the limit where $U/t\to\infty$, the ground state -- which is effectively
the ground state of (\ref{heisenberg}) -- has antiferromagnetic
order.  This is the chiral symmetry breaking state. Our mapping of the Coulomb interaction
at half filling onto this class of models would then seem to answer in the affirmative the question
as to whether a strong enough Coulomb interaction would break chiral symmetry.

Indeed, we have thus argued
that, when the Coulomb interaction is dominant, the ground state is a spin density
wave, which gives rise to a chiral symmetry breaking mass term in the Hamiltonian
of the type (\ref{massterm}) and the chiral symmetry is broken as $U(4)\to U(2)\times U(2)$.
It is also clear from the numerical studies that when the parameter $U/t$
is small, the hopping dominates the Hubbard model and the system is in a metallic phase, which is
the one described by the tight binding model of graphene. In reference \cite{paiva}
the phase transition between these two regimes is estimated to be in the range
\begin{equation}\label{crit}4\leq \frac{U}{t}_{\rm crit.}\leq 5\end{equation}
This is very interesting as a rough estimate of this parameter in graphene, $\frac{U}{t}\sim 2$-$4$ puts
it in a regime that is just sub-critical. It is a tantalizing idea that, if graphene could be
modified in a some way, perhaps mechanically by stretching it, to increase $U/t$, this process
could drive a quantum phase transition which would result in chiral symmetry breaking.

Even more intriguing is a recent Monte Carlo study of the Hubbard model on a honeycomb lattice
which finds an intermediate phase in the region
\begin{equation}
3.5\leq \frac{U}{t}\leq 4.3
\end{equation}
which appears to be an RVB spin liquid \cite{meng}.  What such a phase would mean for graphene has not
been explored yet.
We note that we expect that the Mott insulator phase of graphene is well modeled by the Hubbard model.
beyond the first phase transition, the long-ranged fields of the Coulomb interaction could well
take over and the spin-liquid to metal phase transition might not be accurately described by the
Hubbard model.

In a certain sense, we have answered the question as to why such a strong Coulomb interaction
does not break chiral symmetry in graphene.  Even if the coupling is very strong, the first
pass at the strong coupling ground state has a uniform charge distribution, no charge density
wave and no chiral symmetry breaking.  We then need to rely on a subtle effect from degenerate
perturbation theory to make the system an anti-ferromagnet.  In fact, the energy scale of the
antiferromagnet is $t^2/U$ which is smaller if the coupling $U$ is larger.  Put simply, chiral symmetry
breaking is charge density wave formation and a strong Coulomb interaction favors a homogeneous neutral
state over a charge density wave.  We must then rely on more subtle effects to create a spin density wave
which is charge neutral.

\section{Spinless electrons}
\begin{figure}~\includegraphics[scale=1]{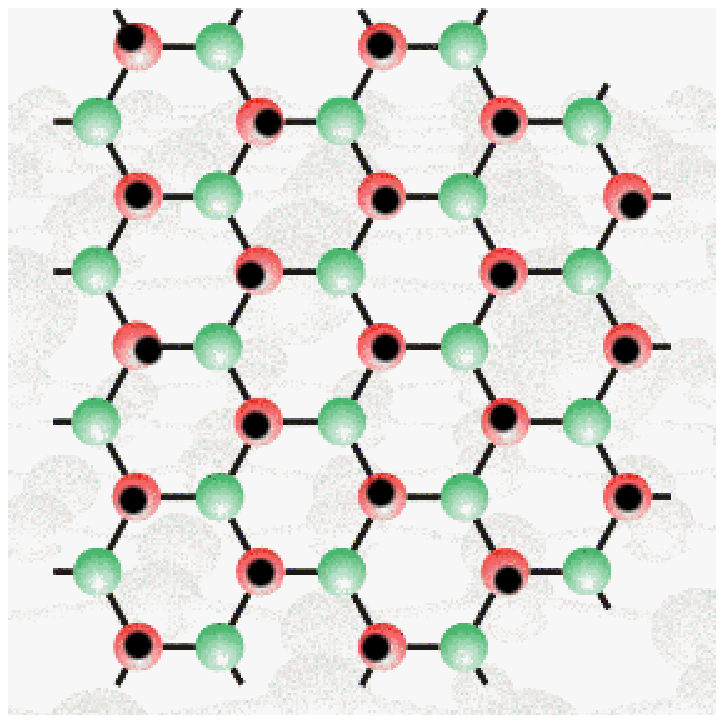}
\begin{caption}{\label{spinpolarized}The charge distribution of the strong coupling ground state of neutral
graphene with a single spin is a Wigner crystal where one of the sublattices is completely
occupied and the other sublattice is completely empty.}
\end{caption}
\end{figure}
To see how results could have been different, consider the case of a half-filled graphene lattice where the
electron has one, rather than two spin states.
In that case, the density of neutral graphene would have one half of the sites of
the lattice occupied.  The -1 in the charge density in (\ref{density}) is replaced by $-\frac{1}{2}$ -- if
the neutral system is to be half-filled, the positive ion residing on each site must have charge $-\frac{1}{2}$.
Thus, the eigenvalues of the charge density at a site are never zero -
the eigenvalues of $\rho_n$ would be $\pm\frac{1}{2}$, depending on whether the
site is occupied or unoccupied.  This means that there is no charge neutral state.
Having no possibility of a neutral state, the electrons do the next best
thing, they maximize the effect of electron-hole attraction, which contributes negatively to the energy
and is strongest if the electron and hole are on nearest neighboring sites.
The Coulomb energy is minimized by the Wigner lattice that is depicted in figure \ref{spinpolarized}. This state
has the same on-site Coulomb energy as any other state (since $\rho_n^2=1/4$) and every set nearest neighbors which have
opposite charges -- and contribute $-\frac{e^2}{4\pi\epsilon a}$ to the energy, where $a$ is the lattice spacing.

Since this state has all of the electrons residing on one of the sublattices, it immediately
breaks chiral symmetry in the maximal possible way.  The hopping term in the Hamiltonian would
have to be very large to restore translation invariance and the metallic state in this system.
Surely, one-spin-state graphene with other parameters similar would be an insulator.

It is interesting that there is such a difference between the two cases, spinful graphene and spinless graphene.  One
would be gapless, the other would be gapped and would be a strong insulator. In the low energy limit,
the only difference between the two is the $U(4)$ versus
$U(2)$ symmetry.  It is an open question as to whether one can discern the difference of critical couplings
at the level of continuum field theory.

\section{Discussion}

\begin{figure}
\includegraphics[scale=.6]{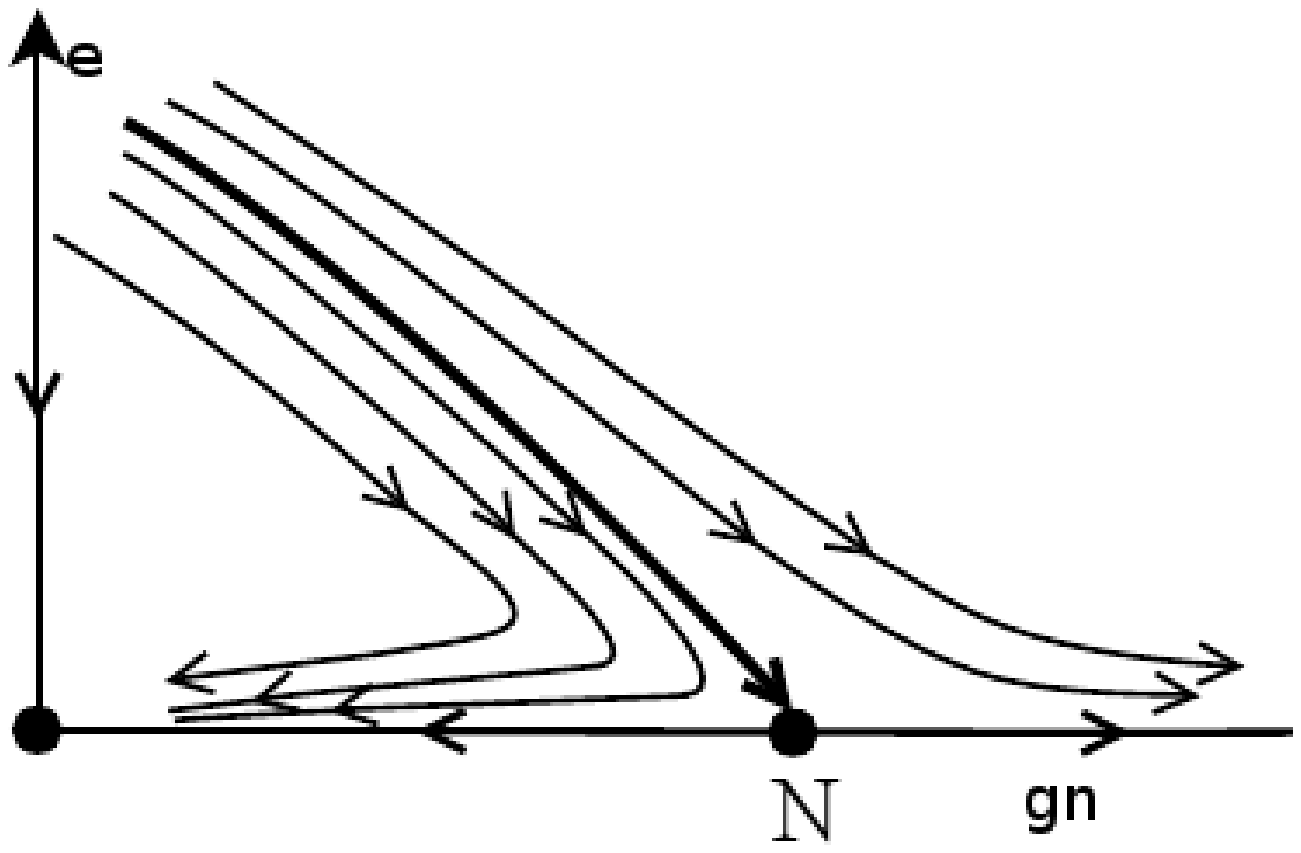}
\begin{caption}{\label{gnfixed} Renormalization group flow for graphene with a Gross-Neveu and Coulomb interaction.}\end{caption}\end{figure}

Coming back to graphene with spin, we observe that, almost for accidental reasons, the most important part of
the Coulomb interaction was the short-ranged on-site interaction.  The relevance of short-ranged part of the electromagnetic interactions   is consistent with renormalization group arguments which
show that local 4-fermi interactions dominate the chiral symmetry
breaking quantum phase transition \cite{juricic}. The renormalization group flow, computed in the $2+\epsilon$ expansion
in reference \cite{juricic} is depicted in figure \ref{gnfixed}. It shows the Gross-Neveu fixed point which dominates the
renormalization group flow when the local point interaction coupling of the Gross-Neveu model (plotted on the horizontal axis) is critical.  When the electromagnetic
interaction (plotted on the vertical axis) is turned on, it lowers the critical coupling somewhat.

\ack{}

The author  acknowledges the financial support of NSERC of Canada
and the kind hospitality of the Aspen Center for Physics,
the KITP Santa Barbara, Galileo Galilei Institute
and Nordita, where parts of this work were completed. This document has been given the KITP
preprint number NSF-KITP-11-183.  Work done at KITP is supported in part by the National
Science Foundation under Grant No.~NSFPHY05-51164 and in part by DARPA under Grant No.
HR0011-09-1-0015 and by the National Science Foundation under Grant
No. PHY05-51164.

\end{document}